\documentclass[longauth]{aa}

\usepackage{url,hyperref}

\usepackage[varg]{txfonts}
\usepackage{graphicx}
\usepackage{natbib}
\bibpunct{(}{)}{;}{a}{}{,} 

\usepackage[switch]{lineno} 

\begin{document}
\title{First limits on the very-high energy gamma-ray afterglow emission of a fast radio burst} 
\subtitle{H.E.S.S. observations of FRB 150418} 

\makeatletter
\renewcommand*{\@fnsymbol}[1]{\ifcase#1\or*\or$\dagger$\or$\ddagger$\or**\or$\dagger\dagger$\or$\ddagger\ddagger$\fi}
\makeatother


\author{\small H.E.S.S. Collaboration
\and H.~Abdalla \inst{1}
\and A.~Abramowski \inst{2}
\and F.~Aharonian \inst{3,4,5}
\and F.~Ait Benkhali \inst{3}
\and A.G.~Akhperjanian\protect\footnotemark[2] \inst{6,5} 
\and T.~Andersson \inst{10}
\and E.O.~Ang\"uner \inst{21}
\and M.~Arakawa \inst{43}
\and M.~Arrieta \inst{15}
% ...
}

\institute{
Centre for Space Research, North-West University, Potchefstroom 2520, South Africa \and 
Universit\"at Hamburg, Institut f\"ur Experimentalphysik, Luruper Chaussee 149, D 22761 Hamburg, Germany \and 
Max-Planck-Institut f\"ur Kernphysik, P.O. Box 103980, D 69029 Heidelberg, Germany \and 
Dublin Institute for Advanced Studies, 31 Fitzwilliam Place, Dublin 2, Ireland \and 
National Academy of Sciences of the Republic of Armenia,  Marshall Baghramian Avenue, 24, 0019 Yerevan, Republic of Armenia  \and
Yerevan Physics Institute, 2 Alikhanian Brothers St., 375036 Yerevan, Armenia \and
Institut f\"ur Physik, Humboldt-Universit\"at zu Berlin, Newtonstr. 15, D 12489 Berlin, Germany \and
University of Namibia, Department of Physics, Private Bag 13301, Windhoek, Namibia \and
GRAPPA, Anton Pannekoek Institute for Astronomy, University of Amsterdam,  Science Park 904, 1098 XH Amsterdam, The Netherlands \and
Department of Physics and Electrical Engineering, Linnaeus University,  351 95 V\"axj\"o, Sweden \and
Institut f\"ur Theoretische Physik, Lehrstuhl IV: Weltraum und Astrophysik, Ruhr-Universit\"at Bochum, D 44780 Bochum, Germany \and
GRAPPA, Anton Pannekoek Institute for Astronomy and Institute of High-Energy Physics, University of Amsterdam,  Science Park 904, 1098 XH Amsterdam, The Netherlands \and
Institut f\"ur Astro- und Teilchenphysik, Leopold-Franzens-Universit\"at Innsbruck, A-6020 Innsbruck, Austria \and
School of Physical Sciences, University of Adelaide, Adelaide 5005, Australia \and
LUTH, Observatoire de Paris, PSL Research University, CNRS, Universit\'e Paris Diderot, 5 Place Jules Janssen, 92190 Meudon, France \and
Sorbonne Universit\'es, UPMC Universit\'e Paris 06, Universit\'e Paris Diderot, Sorbonne Paris Cit\'e, CNRS, Laboratoire de Physique Nucl\'eaire et de Hautes Energies (LPNHE), 4 place Jussieu, F-75252, Paris Cedex 5, France \and
Laboratoire Univers et Particules de Montpellier, Universit\'e Montpellier, CNRS/IN2P3,  CC 72, Place Eug\`ene Bataillon, F-34095 Montpellier Cedex 5, France \and
DSM/Irfu, CEA Saclay, F-91191 Gif-Sur-Yvette Cedex, France \and
Astronomical Observatory, The University of Warsaw, Al. Ujazdowskie 4, 00-478 Warsaw, Poland \and
Aix Marseille Universit\'e, CNRS/IN2P3, CPPM UMR 7346,  13288 Marseille, France \and
Instytut Fizyki J\c{a}drowej PAN, ul. Radzikowskiego 152, 31-342 Krak{\'o}w, Poland \and
Funded by EU FP7 Marie Curie, grant agreement No. PIEF-GA-2012-332350,  \and
School of Physics, University of the Witwatersrand, 1 Jan Smuts Avenue, Braamfontein, Johannesburg, 2050 South Africa \and
Laboratoire d'Annecy-le-Vieux de Physique des Particules, Universit\'{e} Savoie Mont-Blanc, CNRS/IN2P3, F-74941 Annecy-le-Vieux, France \and
Landessternwarte, Universit\"at Heidelberg, K\"onigstuhl, D 69117 Heidelberg, Germany \and
Universit\'e Bordeaux, CNRS/IN2P3, Centre d'\'Etudes Nucl\'eaires de Bordeaux Gradignan, 33175 Gradignan, France \and
Oskar Klein Centre, Department of Physics, Stockholm University, Albanova University Center, SE-10691 Stockholm, Sweden \and
Wallenberg Academy Fellow,  \and
Institut f\"ur Astronomie und Astrophysik, Universit\"at T\"ubingen, Sand 1, D 72076 T\"ubingen, Germany \and
Laboratoire Leprince-Ringuet, Ecole Polytechnique, CNRS/IN2P3, F-91128 Palaiseau, France \and
APC, AstroParticule et Cosmologie, Universit\'{e} Paris Diderot, CNRS/IN2P3, CEA/Irfu, Observatoire de Paris, Sorbonne Paris Cit\'{e}, 10, rue Alice Domon et L\'{e}onie Duquet, 75205 Paris Cedex 13, France \and
Univ. Grenoble Alpes, IPAG,  F-38000 Grenoble, France \protect\\ CNRS, IPAG, F-38000 Grenoble, France \and
Department of Physics and Astronomy, The University of Leicester, University Road, Leicester, LE1 7RH, United Kingdom \and
Nicolaus Copernicus Astronomical Center, Polish Academy of Sciences, ul. Bartycka 18, 00-716 Warsaw, Poland \and
Institut f\"ur Physik und Astronomie, Universit\"at Potsdam,  Karl-Liebknecht-Strasse 24/25, D 14476 Potsdam, Germany \and
Friedrich-Alexander-Universit\"at Erlangen-N\"urnberg, Erlangen Centre for Astroparticle Physics, Erwin-Rommel-Str. 1, D 91058 Erlangen, Germany \and
DESY, D-15738 Zeuthen, Germany \and
Obserwatorium Astronomiczne, Uniwersytet Jagiello{\'n}ski, ul. Orla 171, 30-244 Krak{\'o}w, Poland \and
Centre for Astronomy, Faculty of Physics, Astronomy and Informatics, Nicolaus Copernicus University,  Grudziadzka 5, 87-100 Torun, Poland \and
Department of Physics, University of the Free State,  PO Box 339, Bloemfontein 9300, South Africa \and
Heisenberg Fellow (DFG), ITA Universit\"at Heidelberg, Germany  \and
GRAPPA, Institute of High-Energy Physics, University of Amsterdam,  Science Park 904, 1098 XH Amsterdam, The Netherlands \and
Department of Physics, Rikkyo University, 3-34-1 Nishi-Ikebukuro, Toshima-ku, Tokyo 171-8501, Japan
% ...
}

\offprints{H.E.S.S.~Collaboration,
\\\email{\href{mailto:contact.hess@hess-experiment.eu}{contact.hess@hess-experiment.eu}};
\\$^*$ corresponding authors
\\$\dagger$ Deceased}

\date{14 June 2016}


\abstract {} {Following the detection of the fast radio burst FRB150418 by the SUPERB project at the Parkes radio telescope, we aim to search for very-high energy gamma-ray afterglow emission.} %
{Follow-up observations in the very-high energy gamma-ray domain were obtained with the H.E.S.S. imaging atmospheric Cherenkov telescope system within 14.5 hours of the radio burst.} %
{The obtained 1.4 hours of gamma-ray observations are presented and discussed. At the $99~\%$ C.L. we obtained an integral upper limit on the gamma-ray flux of $\Phi_\gamma(E > 350\,\mathrm{GeV}) < 1.33\times 10^{-8}\,\mathrm{m}^{-2} \mathrm{s}^{-1}$. Differential flux upper limits as function of the photon energy were derived and used to constrain the intrinsic high-energy afterglow emission of FRB 150418.} %
{No hints for high-energy afterglow emission of FRB 150418 were found. Taking absorption on the extragalactic background light into account and assuming a distance of $z=0.492$ based on radio and optical counterpart studies and consistent with the FRB dispersion, we constrain the gamma-ray luminosity at $1\,\mathrm{TeV}$ to $L< 5.1\times 10^{47}\,\mathrm{erg}/\mathrm{s}$ at $99~\%$ C.L.}%
\maketitle

\makeatletter
\renewcommand*{\@fnsymbol}[1]{\ifcase#1\@arabic{#1}\fi}
\makeatother

\section{Introduction}

Fast radio bursts (FRBs) are one of the major astronomical mysteries that have emerged in the last decade. 
First noticed in 2007 in archival data taken with the Parkes radio telescope~\citep{LorimerFRB}, seventeen of these millisecond-duration bursts have been detected so far~\citep{Thornton:2013,Petroff:2015}. The majority were found with the Parkes telescopes, although additional bursts have been detected with the Arecibo telescope~\citep{Spitler:2014} and the Green Bank Telescope (GBT)~\citep{Masui:2015}. 
A summary of known FRBs including the details of the observations can be found in the online catalog FRBCAT\footnote{\tt http://www.astronomy.swin.edu.au/frbcat/} ~\citep{Petroff:2016}.
 
The frequency-dependent dispersion properties of FRBs have constrained their distance to $z\sim0.1-1$~\citep{Petroff:2016}. Distance confusion can, however, arise due to the unknown plasma density within the supposed host galaxy of the FRB, and that of our own Galaxy (the latter is especially relevant in cases where the FRB was observed toward the Galactic plane). 

The typical radio energy output of a few $10^{39}D^2_{\rm 1\,Gpc}$\,erg, assuming isotropic emission at distance $D_{\rm 1\,Gpc}$\;=\;$D/ 1\mathrm{Gpc}$, and the millisecond duration of FRBs have led to proposed scenarios involving compact objects -- white dwarfs (WDs), neutron stars (NSs) and/or black holes (BHs). A review of potential sources can be found for example in~\citet{Kulkarni:2014}.
The merger of various combinations of WDs, NSs and/or BHs are generally favoured (e.g., \citealt{Totani:2013,Zhang:2014,Kashiyama:2013,Mingarelli2015})
in what would be a cataclysmic event similar to short gamma-ray bursts (sGRBs). Other models involve young pulsars created in core-collapse supernovae of massive stars~\citep{Connor01052016} and blitzars~\citep[BH forming rapidly from a NS via accretion,][]{Falcke:2014}. The recent discovery of repeating bursts from FRB121102~\citep{Spitler:2016,Scholz:2016} has renewed attention in non-cataclysmic scenarios such as flares and giant pulses from NSs and/or magnetars~\citep{Lyubarsky:2014,Katz:2015,Pen:2015,Cordes:2016}.

A potentially significant advance in our understanding of FRBs came with the detection of a radio afterglow at the
location of FRB150418 with the Australia Telescope Compact Array~\citep[ATCA,][]{FRB150418_Keane}. The burst FRB150418 was 
initially detected at Parkes on the 18th April 2015 by the SUPERB team. The fading radio afterglow lasted up to 
six days after the FRB, and could be linked to an elliptical host galaxy at $z=0.492\pm0.008$  (WISE\,J071634.59$-$190039.2). 
If connected to the afterglow, the energetics of FRB150418 suggest a cataclysmic origin of the bursts~\citep[e.g.][]{Zhang:2016}. However, alternative explanations for the temporal behavior of the radio flux have been suggested in the form of an unrelated active galactic nucleus (AGN) activity in the host galaxy~\citep{Williams:2016}, or interstellar scintillation~\citep{Akiyama:2016}. Several other possible scenarios could also explain the ATCA source, including an AGN related to the FRB~\citep{Vedantham2016}, a magnetar (so the FRB repeats at the same dispersion measure as FRB150418), localized star formation, a long GRB afterglow~\citep[as seen in GRB130925A,][]{Horesh2015}, or a yet unknown mechanism. Ongoing radio monitoring may resolve the issue in the future.

FRBs release enormous amounts of energy in the radio domain (e.g., FRB150418 released $8^{+1}_{-5} \times 10^{38}\,\mathrm{erg}$ at the position of the potential host galaxy with a luminosity greater than $1.3 \times 10^{42}\,\mathrm{erg/s}$) and their potential origins are thought to be similar to other transients seen in the X-ray and multi-GeV gamma-ray bands such as short and/or long GRBs~\citep{Zhang:2014}. Several FRB models have also specifically suggested the existence of flares in the TeV band (e.g., \citealt{Lyubarsky:2014,Murase:2016}) and proposed follow-ups of FRBs at very high energies.

In this paper, we report the first follow-up observations of FRBs in very high energy (VHE) gamma-rays of TeV (10$^{12}$\,eV) energies. We present observations searching for the very high-energy afterglow of FRB150418 with the High Energy Stereoscopic System (H.E.S.S.) following an alert from the SUPERB collaboration. 


\section{Observations from H.E.S.S. and data analysis}
Dedicated follow-up observations of FRB150418 were obtained in the very-high energy gamma-ray domain with the H.E.S.S. imaging atmospheric Cherenkov telescope array. H.E.S.S. is located on the Khomas Highland plateau of Namibia (23$^{\circ}16'18''$ South, $16^{\circ}30'00''$ East), at an elevation of 1800 m above sea level. With its original four-telescope array, H.E.S.S. is sensitive to cosmic and gamma-rays in the 100\,GeV to 100\,TeV energy range and is capable of detecting a Crab-like source close to zenith and under good observational conditions at the 5$\sigma$ level within less than one minute~\citep{HESS-Crab2006}. In 2012 a fifth telescope with 28\,m diameter was commissioned, extending the covered energy range toward lower energies. This fifth telescope was unavailable at the time of the observation and data for the follow-up presented here have therefore been obtained with the four 12\,m H.E.S.S. telescopes.

\begin{figure}[!h]
\centering
\includegraphics[width=0.48\textwidth]{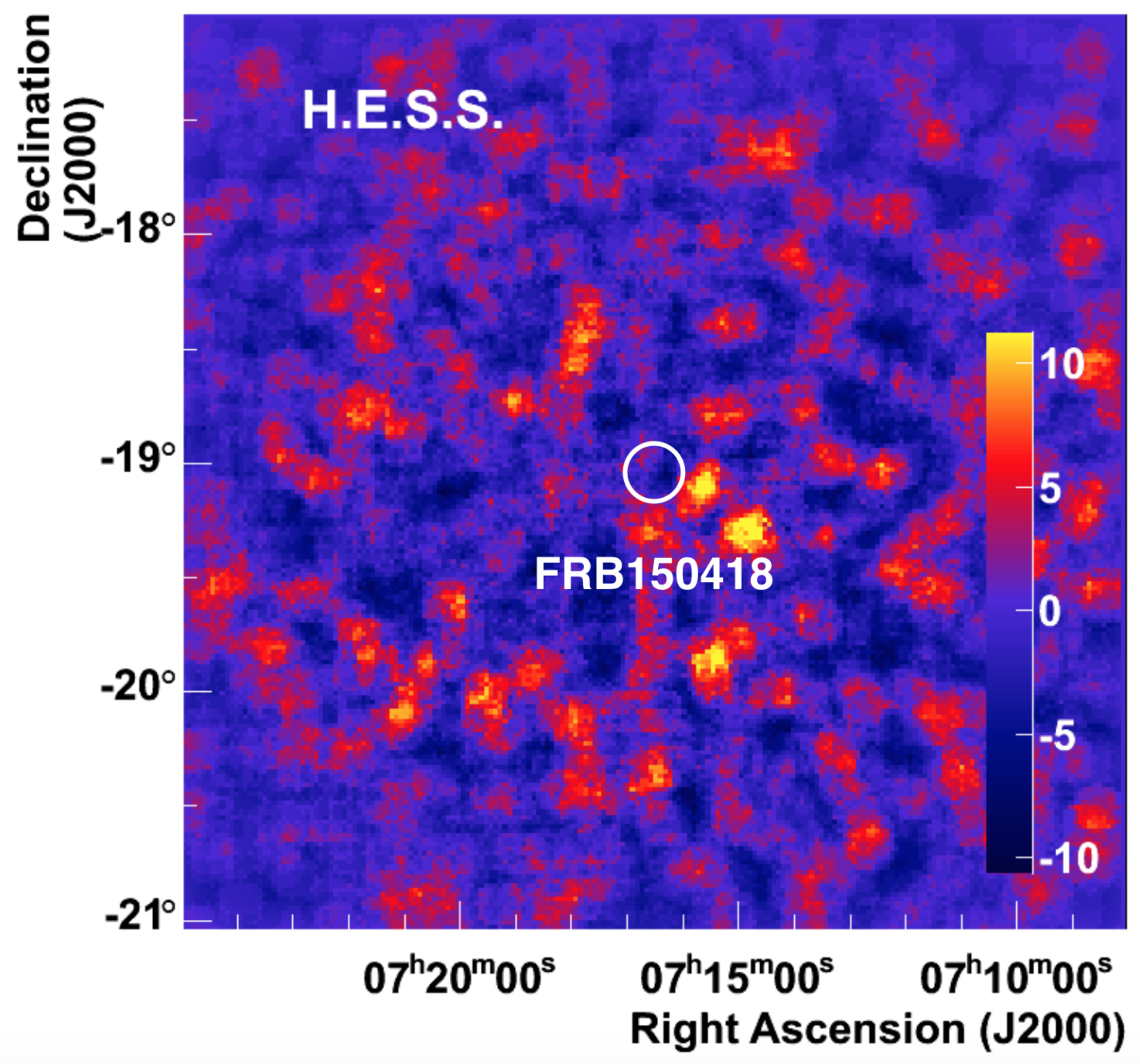}
\caption{VHE gamma-ray emission around the direction of FRB150418 illustrated by the event counts exceeding the background. The circle in the center has a diameter of $0.24^\circ$ and denotes the width of the Parkes beam in which the burst has been observed.}\label{fig:excess}
\end{figure}

\begin{figure*}[!th]
  \resizebox{\hsize}{!}{
  \includegraphics[width=0.48\textwidth]{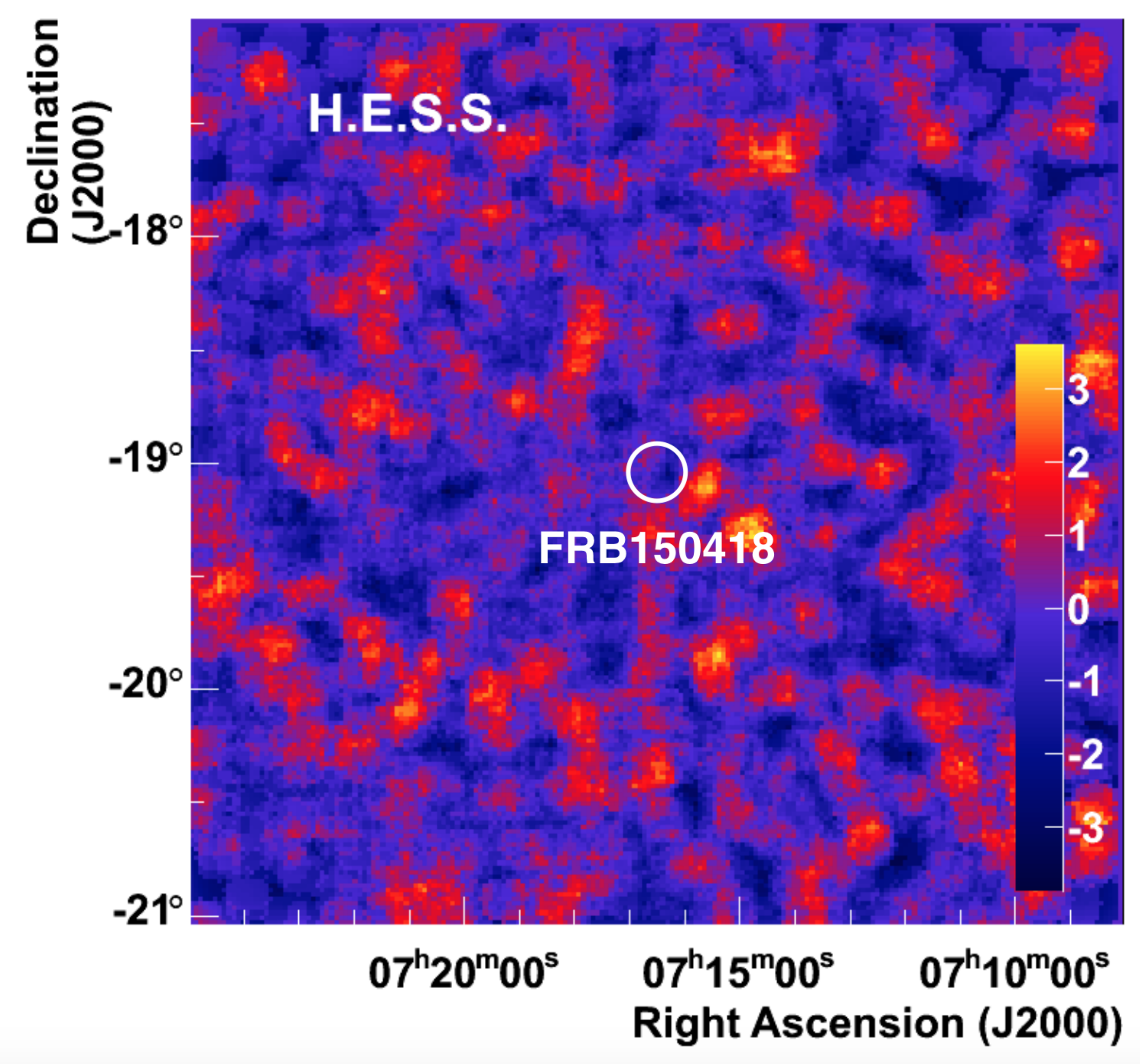}%
  \includegraphics[width=0.48\textwidth]{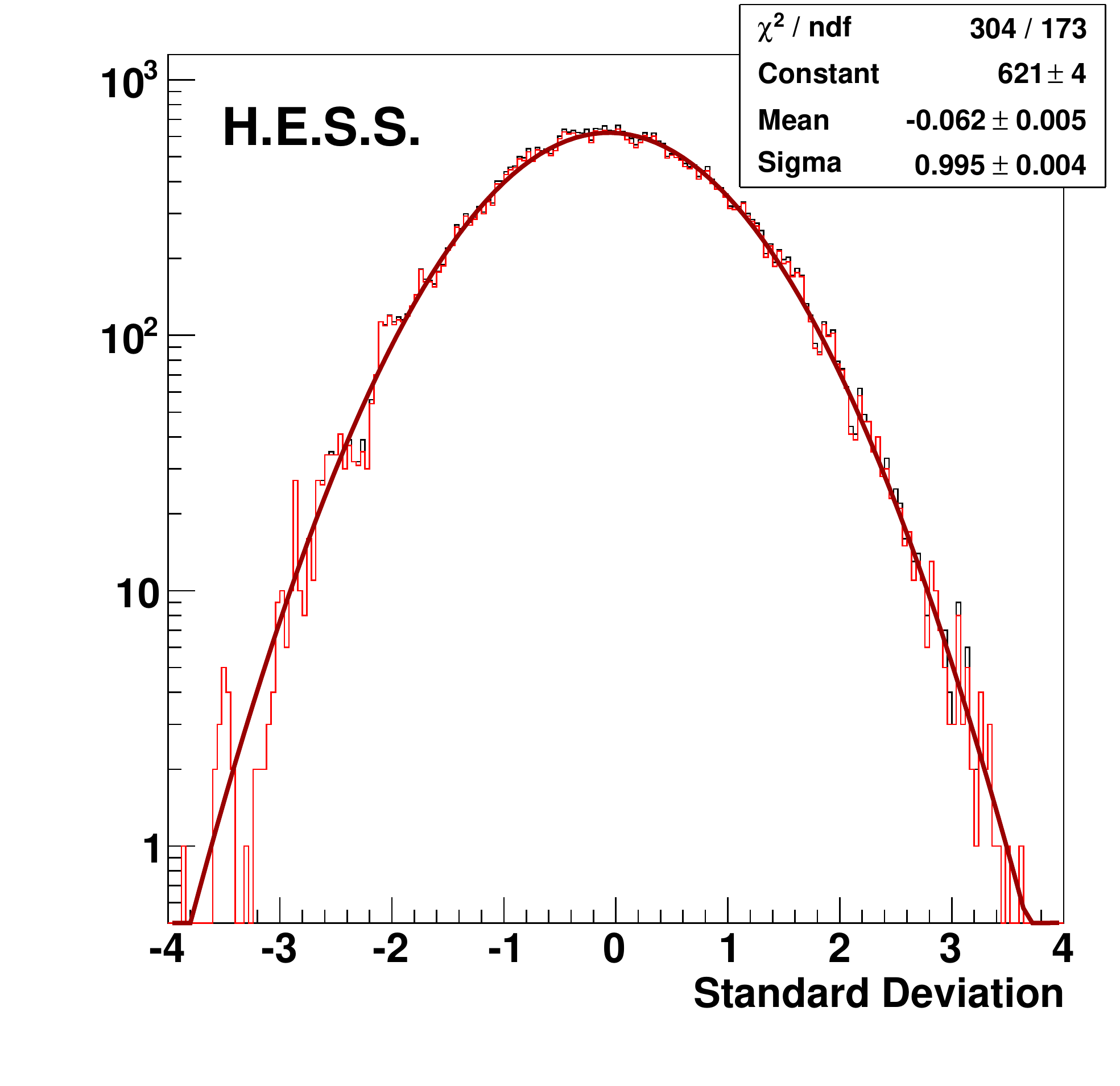}
  }
\caption{Left plot: Map of significances of the gamma ray emission using the formalism proposed by~\citet{LiMa} in the region around FRB150418. The circle in the center has a diameter of $0.24^\circ$ and denotes the width of the Parkes beam in which the burst was observed. Right plot: Distribution of significances (black histogram) compared to the distribution obtained by excluding a circular region of $0.25^\circ$ radius (red histogram). The red line and the shown parameters correspond to a Gaussian function fit to the latter distribution.}\label{fig:Significance}
\end{figure*}

The notification of FRB150418 was received from the SUPERB team on 2015-04-18 during daytime at the site of the H.E.S.S. experiment, thus prohibiting prompt follow-up observations. The necessary observation conditions were reached the evening of the same day at 17:55 UTC (about 14.5\,h after the FRB) and 1.4\,h of data could be recorded until the source set below an elevation of $45^\circ$, which is the typical horizon for observations retaining a relatively low energy threshold. The data, taken in standard wobble mode operations with source offsets of $0.7^\circ$, fulfill all standard data quality criteria including requirements on atmospheric conditions, and detector stability. The zenith angle of the observations ranged from $21^\circ$ to $42^\circ$. After correcting for acceptance effects due to the wobble source offsets, a total effective live-time of 1.1\,h at the FRB position was available for analysis.

The data were analyzed using Model Analysis~\citep{ModelAnalysis}, an advanced Cherenkov image reconstruction method in which the recorded shower images of all triggered telescopes are compared to a semi-analytical model of gamma ray showers by means of a log-likelihood optimization. The ``standard cuts'' of Model Analysis were adopted. These cuts require, among other criteria, the total charge in the shower image to be greater than 60 photoelectrons. The resulting energy threshold, defined as the energy where the acceptance is $20\%$ of its maximum value, is 350\,GeV for this dataset. 

The robustness and stability of the described analysis have been verified with an independent analysis relying on an independent data calibration chain and using the Image Pixel-wise fit for Atmospheric Cherenkov Telescopes~\citep[ImPACT,][]{ImPACT} reconstruction method. The results from this cross-check analysis are consistent with the ones presented here.

The H.E.S.S. field-of-view (FoV) with a diameter of $5^\circ$ easily covers the Parkes beam with a FWHM of $0.24^\circ$~\citep{FRB150418_Keane}. The H.E.S.S. observations therefore cover all potential locations of FRB150418 within the Parkes beam in which the FRB was detected. On the other hand, the H.E.S.S. point-spread function has a diameter of $\sim\!0.12^\circ$ ($68~\%$ containment), that is, half the Parkes beam size. We can therefore, based solely on H.E.S.S. data, not expect to easily resolve the origin of a potential afterglow within the Parkes beam and would not be able to discriminate between the potential host galaxy discussed by~\cite{FRB150418_Keane} and other locations within the beam.

No high-energy gamma ray source has been detected within the region of interest in the four-year long observations by the LAT instrument onboard the Fermi satellite~\citep{3FGL}. Also no emission at very-high gamma-ray energies has been reported so far from the region~\footnote{\tt http://tevcat.uchicago.edu}. 

The background level in the FoV was determined from the dataset itself using the standard ``ring background'' technique~\citep{RingBg}, a robust method ideally suited to deriving gamma-ray emission maps in FoVs with low numbers of sources. In order to derive the acceptance function required as input to the ring background method we exploited the azimuthal symmetry of the acceptance across the field-of-view of the telescopes. We derive the acceptance from the same dataset and, in order to reduce systematic uncertainties due to the limited statistics, we refrained from a detailed modeling of the zenith angle dependence of the acceptance function and use the acceptance derived at the average zenith angle of $32^\circ$.

%


\section{Results}

The map of gamma-ray events exceeding the background is shown for the full region of interest (ROI) around FRB150418 in Fig.~\ref{fig:excess}. We then converted the excess counts into significance levels using the formalism described by~\cite{LiMa}. The resulting map of significances is shown in the left plot of Fig.~\ref{fig:Significance}. It should be noted that trial factors due to the large number of individual bins are not accounted for in this representation. For an ROI dominated by statistical fluctuations of the background the distribution of the significances should follow a Gaussian with a mean at zero and a width of one. The right plot in Fig.~\ref{fig:Significance} shows the corresponding distribution (black histogram). The distribution obtained by excluding a circular "signal" region of $0.25^\circ$ radius around the FRB position is shown in red. Both histograms agree very well. In addition, when fitting the latter distribution with a Gaussian shape, very good agreement with the ``background only'' hypothesis was found. It can be noted that the errors on the obtained parameters are underestimated due to correlations in the entries of the significance distributions which are introduced by the background estimation on overlapping regions. We conclude that the ROI is well described and clearly dominated by background events.

\begin{figure}[!th]
\centering
\includegraphics[width=0.48\textwidth]{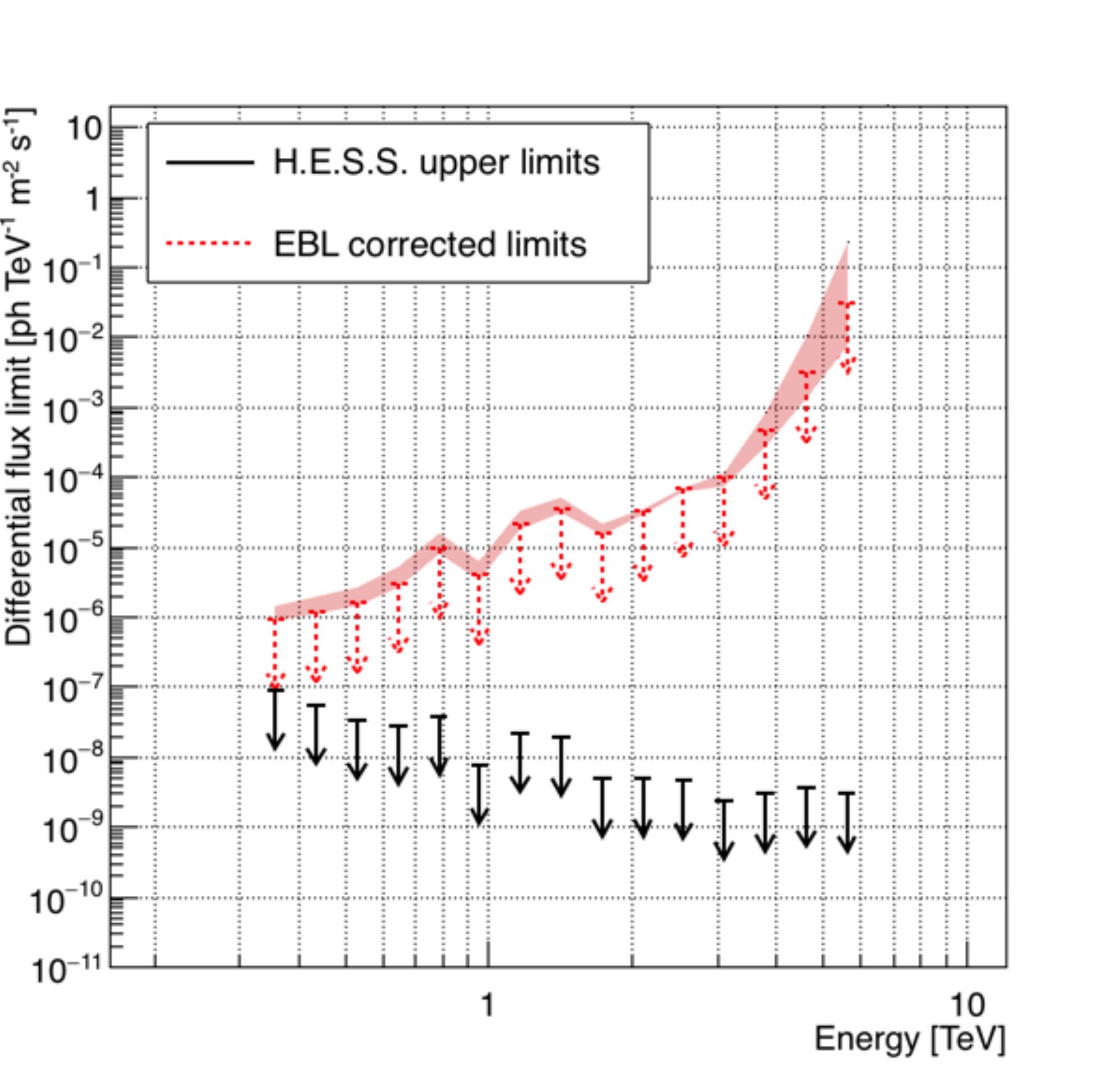}
\caption{Limits (99\,\% CL) on the very high energy gamma-ray flux derived from the H.E.S.S. follow-up on FRB150418 assuming an $E^{-2}$ energy spectrum. The EBL de-absorption is based on the model from~\citet{Gilmore:2011ks} and assumes the FRB distance of $z=0.492$~\citep{FRB150418_Keane}. The uncertainty induced by different EBL models is shown as red band.}\label{fig:UL}
\end{figure}

As the obtained results were fully compatible with the background expectation we conclude that no significant gamma-ray afterglow was detected from the direction of FRB150418 (c.f. Fig.~\ref{fig:Significance}). Consequently we derive $99\,\%$ C.L. upper limits on the gamma-ray flux as function of energy following the approach by~\citet{Feldman1998}. Assuming a generic $E^{-2}$ energy spectrum for the potential emission and integrating above the threshold of $350\,\mathrm{GeV}$ we obtain $\Phi_\gamma(E > 350\,\mathrm{GeV}) < 1.33\times 10^{-8}\,\mathrm{m}^{-2} \mathrm{s}^{-1}$. Assuming a $E^{-4}$ energy spectrum, we obtain $\Phi_\gamma(E > 350\,\mathrm{GeV}) < 2.12\times 10^{-8}\,\mathrm{m}^{-2} \mathrm{s}^{-1}$. 
Differential upper limits as function of the energy are shown as black arrows in Fig.~\ref{fig:UL}. Due to the small size of the bins, the influence of the assumed spectrum (e.g. $E^{-2}$ vs. $E^{-4}$) on the differential upper limits is less than $1.3~\%$.

While propagating through the extragalactic radiation fields, high-energy gamma rays interact with the extragalactic background light (EBL) via $e^+ / e^-$ pair-creation processes. This leads to the collective effect of an absorption of gamma-rays at the highest energies. The resulting gamma-ray opacity depends on the energy as well as on the distance of the source. Using the EBL model published in~\cite{Gilmore:2011ks} we were able to correct the derived upper limits on the gamma-ray flux measured on Earth for EBL absorption effects and thus derive energy dependent intrinsic flux limits of the FRB. The result shown in Fig.~\ref{fig:UL} has been derived using the redshift of the potential host galaxy of FRB150418, $z=0.492$~\citep{FRB150418_Keane}. While this distance is consistent with the one derived from the dispersion measure of the FRB, there is still controversy as to the relationship between FRB150418 and WISE\,J071634.59$-$190039.2. The intrinsic limits shown should therefore not be taken as definitive, but rather as an illustration of how the EBL absorption impacts the constraints as a function of energy. The red band in Fig.~\ref{fig:UL} illustrates the effect of different EBL models (e.g. ~\citealt{Gilmore:2011ks, Franceschini2008, Dominguez:2010bv}) on the EBL correction of the derived flux limits.

\section{Discussion and conclusion}
We have reported the first follow-up observations of fast radio bursts in the very high-energy gamma-ray domain. The origin of FRBs remains elusive, and observational constraints such as those presented here are crucial pieces for solving this puzzle. In addition to an enlarged wavelength coverage, timely observations are essential in order to be able to cover as many of the potentially very rapid emission scenarios as possible. 

The luminosity in the radio domain of FRB150408 has been estimated to $L>1.3 \times 10^{42}\,\mathrm{erg}/\mathrm{s}$~\citep{FRB150418_Keane}. The first non-radio observations of the emission region of FRB150418 were carried out 8h after the radio burst by the Swift X-ray satellite and a $3\sigma$ upper limit on the X-ray flux of $\Phi_\mathrm{X} < 7.1 \times 10^{-14}\,\mathrm{erg}\,\mathrm{cm}^{-2}\,\mathrm{s}^{-1}$ has been derived~\citep{FRB150418_Keane}. Our VHE gamma-ray observations constrain emission at slightly longer timescales (starting 14.5\,h after the burst, due to the inability of Imaging Atmospheric Cherenkov Telescopes to observe during daytime) and provide valuable input to models specifically suggesting flares in the TeV band~\citep{Lyubarsky:2014,Murase:2016}. Taking absorption on the extragalactic background light into account, as shown in Fig.~\ref{fig:UL}, and assuming a distance of $z=0.492$ based on radio and optical counterpart studies and consistent with the FRB dispersion, we constrain the gamma-ray luminosity of the afterglow of FRB150418 at $1\,\mathrm{TeV}$ to $L< 5.1\times 10^{47}\,\mathrm{erg}/\mathrm{s}$ at $99\,\%$ C.L. 

\begin{acknowledgements}
The support of the Namibian authorities and of the University of Namibia in facilitating the construction and operation of H.E.S.S. is gratefully acknowledged, as is the support by the German Ministry for Education and Research (BMBF), the Max Planck Society, the German Research Foundation (DFG), the French Ministry for Research, the CNRS-IN2P3 and the Astroparticle Interdisciplinary Programme of the CNRS, the U.K. Science and Technology Facilities Council (STFC), the IPNP of the Charles University, the Czech Science Foundation, the Polish Ministry of Science and Higher Education, the South African Department of Science and Technology and National Research Foundation, the University of Namibia, the Innsbruck University, the Austrian Science Fund (FWF), and the Austrian Federal Ministry for Science, Research and Economy, and by the University of Adelaide and the Australian Research Council. We appreciate the excellent work of the technical support staff in Berlin, Durham, Hamburg, Heidelberg, Palaiseau, Paris, Saclay, and in Namibia in the construction and operation of the equipment. This work benefited from services provided by the H.E.S.S. Virtual Organisation, supported by the national resource providers of the EGI Federation.
\end{acknowledgements}
\bibliographystyle{aa} 
\bibliography{FRB150418}

\end{document}